\documentclass[epjST]{svjour}
\usepackage{graphics}
\usepackage{amsmath}
\usepackage{amssymb}

\newcommand{\be}{\begin{equation}}
\newcommand{\ee}{\end{equation}}
\newcommand{\MS}{\overline{\mathrm{MS}}}
\newcommand{\MC}{\mathrm{MC}}
\newcommand{\NP}{\mathrm{NP}}
\newcommand{\G}{\mathrm{G}}
\newcommand{\latt}{\mathrm{latt}}
\newcommand{\nn}{\nonumber}
\newcommand{\lQ}{\Lambda_{\mathrm{QCD}}}
\newcommand{\al}{\alpha}

\newcommand{\bea}{\begin{eqnarray}}
\newcommand{\eea}{\end{eqnarray}}

\begin{document}
\title{Theoretical description of the plaquette with exponential accuracy}
\author{Antonio Pineda\inst{1,2}\fnmsep\thanks{\email{pineda@ifae.es}}}
\institute{ Institut de F\'\i sica d'Altes Energies (IFAE), \\
The Barcelona Institute of Science and Technology, \\
Campus UAB, 08193 Bellaterra (Barcelona), Spain\and
Grup de F\'{\i}sica Te\`orica, Dept. F\'\i sica, \\
Universitat Aut\`onoma de Barcelona,
E-08193 Bellaterra, Barcelona, Spain\\
}
\abstract{
We review recent studies of the operator product expansion of the plaquette and of the associated determination of the gluon condensate. One first needs the perturbative expansion to orders high enough to reach the asymptotic regime where the renormalon behavior sets in. The divergent perturbative series is formally regulated using the principal value prescription for its Borel integral.  Subtracting the perturbative series truncated at the minimal term, we obtain 
 the leading non-perturbative correction of the operator product expansion, i.e., the gluon condensate, with superasymptotic accuracy. It is then explored how to increase such precision within the context of the hyperasymptotic expansion. The results fully confirm expectations from renormalons and the operator product expansion.
} %end of abstract
\maketitle
\section{Introduction}

The operator product expansion (OPE)~\cite{Wilson:1969zs} is a fundamental
tool for theoretical analyses in quantum field theories.
Its validity is only proven
rigorously within perturbation theory to arbitrary finite
orders~\cite{Zimmermann:1972tv}. The use of the OPE
in a non-perturbative framework was initiated
by the ITEP group~\cite{Vainshtein:1978wd}
(see also the discussion in~Ref.~\cite{Novikov:1984rf}), who postulated
that the OPE of a correlator could be approximated by the following series:
\be
\label{eq:ope1}
\mathrm{correlator}(Q) \simeq \sum_d\frac{1}{Q^d}C_d(\alpha)
\langle O_d \rangle
\,,
\ee  
where the expectation values of local operators $O_d$
are suppressed by inverse powers of a large
external momentum $Q\gg\lQ$, according to their dimensionality $d$.
The Wilson coefficients $C_d(\alpha)$ 
encode the physics at momentum scales larger than $Q$.
These are well approximated by perturbative
expansions in the strong coupling parameter $\alpha$:
\be
C_d(\alpha) \simeq \sum_{n\geq 0} c_n\al^{n+1}\,.
\ee
The large-distance
physics is described by the matrix elements
$\langle O_d \rangle$ that usually have to be
determined non-perturbatively: $\langle O_d \rangle \sim \lQ^d$.

It can hardly be overemphasized that (except for direct
predictions of non-perturbative lattice simulations, e.g., on
light hadron masses)
all QCD predictions are based on factorizations that are generalizations of the above generic OPE.

In this short review, we summarize and discuss the
recent results~\cite{Bali:2014fea,Bali:2014sja,Ayala:2019uaw,Ayala:2019hkn,Ayala:2019lak,Ayala:2020pxq} (see also the review \cite{Bali:2015cxa}), which 
 validate the nonperturbative version of the OPE for the case of the plaquette in gluodynamics.
This analysis utilizes lattice regularization. Then main advantage of this choice is that it enables
us to use numerical stochastic perturbation
theory (NSPT)~\cite{DRMMOLatt94,DRMMO94,DR0} to obtain perturbative expansion
coefficients. This allows us to realize much higher orders than
would have been possible with diagrammatic techniques.
A disadvantage of the lattice scheme is that, at least in our discretization,
lattice perturbative expansions converge slower than expansions in
the $\MS$ coupling. This means that we have to go to comparatively higher
orders to become sensitive to the asymptotic behavior. Many of the results
obtained in a lattice scheme either directly apply to the $\MS$
scheme too or can easily (and in some cases exactly)
be converted into this scheme. We also show how to obtain a theoretical controlled expression for the plaquette with exponential accuracy that is, in principle, systematically improbable using hyperasymptotic expansions, as developed in \cite{Ayala:2019uaw,Ayala:2019hkn,Ayala:2019lak,Ayala:2020pxq} (see \cite{BerryandHowls,Boyd99} for original work in the context of ordinary differential equations).

The expectation value
of the plaquette calculated in Monte Carlo
(MC) simulations in lattice regularization with the
standard Wilson gauge action~\cite{Wilson:1974sk} reads
\begin{equation}
\langle P\rangle_{\mathrm{MC}}=\frac{1}{N^4}\sum_{x\in\Lambda_E}\langle P_x\rangle\,,
\end{equation}
where $\Lambda_E$ is a Euclidean spacetime lattice and
\begin{equation}
P_{x,\mu\nu}=1-\frac{1}{6}\mathrm{Tr}\left(U_{x,\mu\nu}+U_{x,\mu\nu}^{\dagger}\right)\,.
\end{equation}
For details on
the notation, see Ref.~\cite{Bali:2014fea}. 

\section{The plaquette: OPE in perturbation theory}

$\langle P\rangle$ depends on the lattice extent $Na$, the spacing $a$ and
$\al =g^2/(4\pi) \equiv \al(a^{-1})$ (note that $\al$ is the bare lattice coupling 
and its natural scale is of order $a^{-1}$). To compute this expectation value in strict perturbation theory, we Taylor expand in powers of $g$ {\it before} averaging over the gauge configurations (which we do using NSPT~\cite{DRMMOLatt94,DRMMO94,DR0}). 
The outcome is a power series in $\al$:
\begin{equation}
\nn
\langle P \rangle_{\mathrm{pert}}(N) \equiv \frac{1}{Z}\left.\int\![dU_{x,\mu}]\,e^{-S[U]} P[U]\right|_{\mathrm{NSPT}}
=\sum_{n\geq 0}p_n(N)\al^{n+1}\,.
\end{equation}
The dimensionless coefficients $p_n(N)$ are functions of the linear
lattice size $N$. We emphasize that they
do not depend on the lattice spacing $a$, nor on the physical
lattice extent $Na$, alone, but only on
the ratio $N=(Na)/a$.

We are interested in the large-$N$ (i.e., infinite volume)
limit. In this situation 
\be
\label{eq:scales}
\frac{1}{a} \gg \frac{1}{Na}
\ee
and it makes sense to factorize the contributions of the
different scales within the OPE
framework (in perturbation theory). 
The hard modes, of scale $\sim 1/a$,
determine the Wilson
coefficients, whereas the soft modes, of scale $\sim 1/(Na)$, can be described
by expectation values of local gauge invariant operators. There are no
such operators of dimension two.
The renormalization group invariant definition of the gluon condensate
\be
\label{eq:GC}
\langle G^2 \rangle=-\frac{2}{\beta_0}\left\langle\Omega\left| \frac{\beta(\alpha)}{\alpha}
G_{\mu\nu}^cG_{\mu\nu}^c\right|\Omega\right\rangle
=
\left\langle\Omega\left|  \left[1+\mathcal{O}(\alpha)\right]\frac{\alpha}{\pi}
G_{\mu\nu}^cG_{\mu\nu}^c\right|\Omega\right\rangle
\ee
is the only local gauge invariant expectation value of an operator of dimension
$a^{-4}$ in pure gluodynamics. In the purely perturbative case discussed here,
it only depends on the soft scale $1/(Na)$, i.e.\ on the lattice
extent. 
On dimensional grounds, the perturbative gluon condensate
$\langle G^2 \rangle_{\mathrm{soft}}$ is proportional to 
$1/(Na)^4$, and the logarithmic $(Na)$-dependence is encoded
in  $\al[1/(Na)]$. Therefore, 
\be
\label{eq:fnnn}
\frac{\pi^2}{36}\,a^4\langle G^2\rangle_{\mathrm{soft}}=
-\frac{1}{N^4}
\sum_{n\geq 0}f_n\al^{n+1}\![1/(Na)]\,,
\ee
and the perturbative expansion of the plaquette on a finite
volume of $N^4$ sites can be written as
\be
\label{OPEpert}
\langle P \rangle_{\mathrm{pert}} (N)=
P_{\mathrm{pert}}(\al)\langle 1 \rangle
+\frac{\pi^2}{36}C_{\G}(\al)\,a^4\langle G^2\rangle_{\mathrm{soft}}
+\mathcal{O}\left(\frac{1}{N^6}\right)\,,
\ee
where 
\be
P_{\mathrm{pert}}(\al)=\sum_{n\geq 0}p_n\al^{n+1}
\ee
and $p_n$ are the infinite volume coefficients that we
are interested in. The Wilson coefficient, $C_{\G}$,
which only depends on $\al$, 
is normalized to unity for $\al=0$. It can be expanded in $\al$:
\be
\label{CG}
C_{\G}(\al)=1+\sum_{k\geq 0}c_k\al^{k+1}
\,.
\ee
Since the Wilson action is proportional to the plaquette $P$, $C_{\G}$ is
fixed by the conformal trace
anomaly~\cite{DiGiacomo:1990gy,DiGiacomo:1989id}:
\be
\label{CP}
C^{-1}_{\G}(\al)=
-\frac{2\pi\beta(\al)}{\beta_0\al^2}
=1+\frac{\beta_1}{\beta_0}\frac{\al}{4\pi}
+\frac{\beta_2}{\beta_0}\left(\frac{\al}{4\pi}\right)^2
+\frac{\beta_3}{\beta_0}\left(\frac{\al}{4\pi}\right)^3
+\mathcal{O}(\al^4)\,.
\ee
The $\beta$-function coefficients\footnote{We define the $\beta$-function as $\beta(\al)=d\al/d\ln\mu=-\beta_0/(2\pi)\alpha^2-\beta_1/(8\pi^2)\al^3-\cdots$, i.e.\ $\beta_0=11$.} $\beta_j$  are known in the Wilson action lattice
scheme for $j\leq 3$ (see Eq.~(25) of Ref.~\cite{Bali:2014fea}). $\beta_2^{\latt}$ has been
computed diagrammatically~\cite{Luscher:1995np,Christou:1998ws,Bode:2001uz}.
The value for $\beta_3^{\latt}$ that we use~\cite{Bali:2013qla} is an update of \cite{Bali:2013pla}, 
and was obtained by calculating the
normalization of the leading renormalon of the pole mass, and
then assuming the corresponding $\MS$-scheme 
expansion to follow its asymptotic behaviour
from orders $\alpha^4$ onwards. 
Similar estimates, $\beta_3^{\latt}\approx
-1.37\times 10^6$ up to $\beta_3^{\latt}\approx
-1.55\times 10^6$, were found in Ref.~\cite{Guagnelli:2002ia}
using a very different method. Note that $C_{\rm G}(\al)$ is
scheme-dependent not only through $\alpha$, but also
explicitly, due to its dependence on the higher $\beta$-function
coefficients: $\beta_2$, etc..
The $c_k$ 
depend on the $\beta_i$ with $i\leq k+1$ via \eqref{CP}.
For $j\leq 3$ the coefficients $\beta_j$  are known in the Wilson action lattice
scheme.  For convenience, we also write the expansion coefficients $c_k$
defined in \eqref{CP} in terms of the constants that appear in
\be
\label{eq:betafun}
\Lambda=\mu\exp\left\{-\left[\frac{2\pi}{\beta_0\alpha(\mu)}
+b
\ln\left(\frac12 \frac{\beta_0\alpha(\mu)}{2\pi}\right)
+\sum_{j\geq 1}
s_j\,(-b)^j\!\left(\frac{\beta_0\alpha(\mu)}{2\pi}\right)^{\!j}\right]\right\}
\ee
where
$b=\beta_1/(2\beta_0^2)$,
$s_1=(\beta_1^2-\beta_0\beta_2)/(4b\beta_0^4)$ and
$s_2=(\beta_1^3-2\beta_0\beta_1\beta_2+\beta_0^2\beta_3)/(16b^2\beta_0^6)$
and
\begin{equation}
\label{eq:relatecs}
c_0=-b\frac{\beta_0}{2\pi}\,,\quad
c_1=s_1b\left(\frac{\beta_0}{2\pi}\right)^{\!2}\,,\quad
c_2=-2s_2b^2\left(\frac{\beta_0}{2\pi}\right)^{\!3}\,.
\end{equation}

\begin{figure}[t]
\begin{center}
\resizebox{0.75\columnwidth}{!}{  \includegraphics{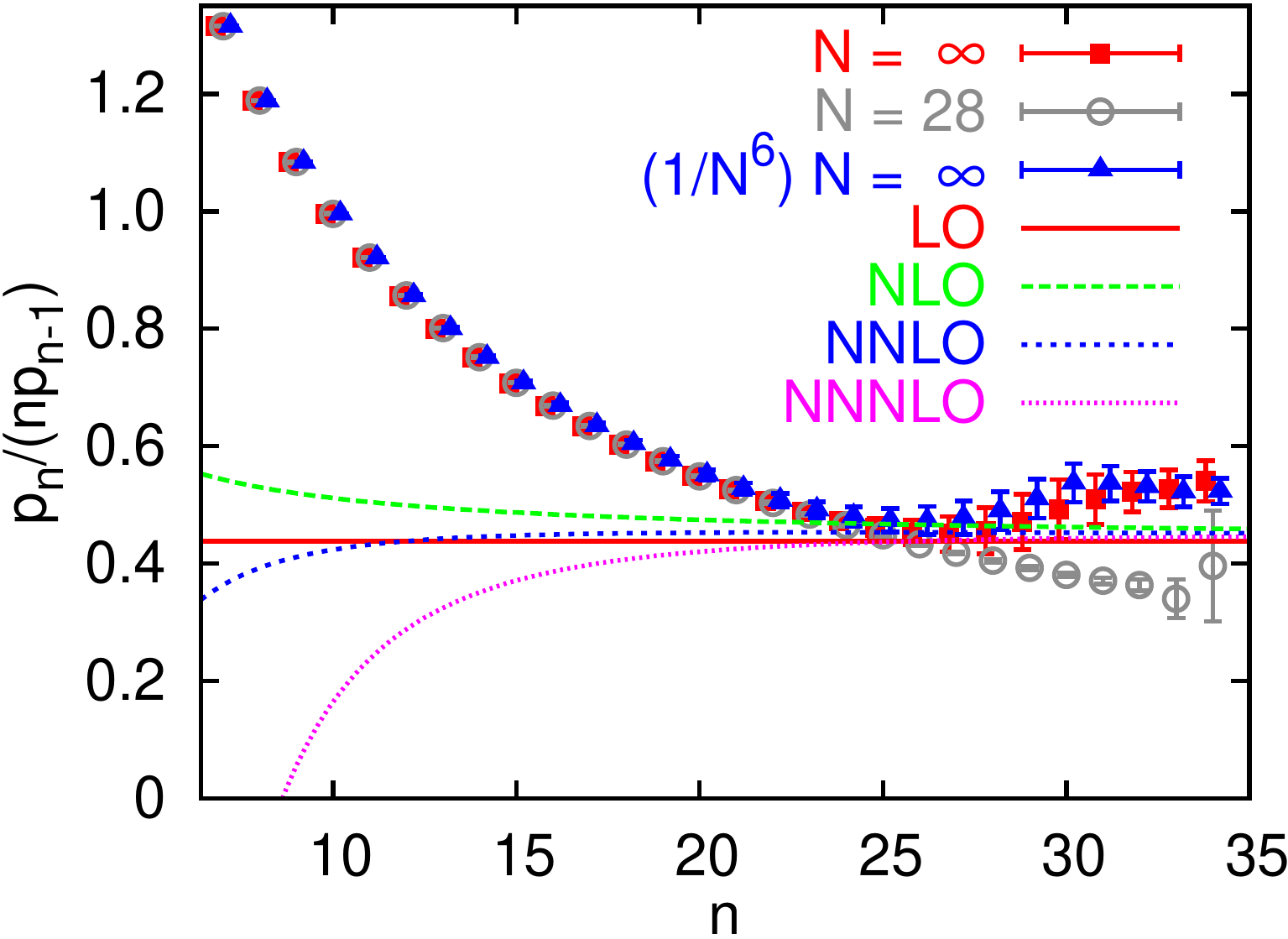} }
\end{center}
\caption{{The ratios
$p_n/(np_{n-1})$ compared with the leading order (LO),
next-to-leading order
(NLO), NNLO and NNNLO predictions of the
$1/n$-expansion Eq. \ref{th:ratio}.
Only the ``$N=\infty$'' extrapolation includes the systematic uncertainties.
We also show finite volume data for $N=28$, and the result from the
alternative $N\rightarrow\infty$ extrapolation including some $1/N^6$
corrections. The symbols have been shifted slightly horizontally. From \cite{Bali:2014fea}.}
\label{n35}}
\end{figure}

Combining Eqs. (\ref{eq:fnnn}), (\ref{OPEpert}) and (\ref{CG})  gives 
\bea
\label{PpertOPE}
\langle P \rangle_{\mathrm{pert}}(N)
&=&\sum_{n\geq 0}\left[p_n-\frac{f_n(N)}{N^4}\right]\al^{n+1}\\\nn
&=&
\sum_{n\geq 0}p_n\al^{n+1}
-\frac{1}{N^4}
\left(1+\sum_{k\geq 0}c_k\al^{k+1}(a^{-1})
\right)
\times 
\sum_{n\geq 0}f_n\al^{n+1}((Na)^{-1})
+\mathcal{O}\left(\frac{1}{N^6}\right)\,,
\eea
where $f_n(N)$ is a polynomial in powers of $\ln(N)$. 
Fitting this equation to the perturbative lattice results,
the first 35 coefficients $p_n$ were determined
in Ref.~\cite{Bali:2014fea}. The results were confronted
with the expectations from renormalons:
\be
\label{eq:thratio1}
p^{\latt}_n \stackrel{n\rightarrow\infty}{=}
Z^{\latt}_{P}\,\left(\frac{\beta_0}{2\pi d}\right)^{\!n}
\frac{\Gamma(n+1+db)}{\Gamma(1+db)} \left[
1+\frac{20.08931\ldots}{n+db}+\frac{505\pm 33}{\left(n+db\right) \left(n+db-1\right)}
+
\mathcal{O}\left(\frac{1}{n^3}\right)
\right]
\,,
\ee
\be
\label{th:ratio}
\frac{p_n}{np_{n-1}}=\frac{\beta_0}{2\pi d}
\left\{1
+\frac{db}{n}
+\frac{db(1-ds_1)}{n^2}
+\frac{db\left[1-3ds_1+d^2b(s_1+2s_2)\right]}{n^3}+
\mathcal{O}\left(\frac{1}{n^4}\right)
\right\}\,.
\ee

In Fig.~\ref{n35}, the infinite volume ratios
$p_n/(np_{n-1})$ are compared to the expectation Eq. \ref{th:ratio}. 
The asymptotic behavior of the perturbative series due to
renormalons is reached around orders $n \sim 27-30$, proving,
for the first time, the existence of the renormalon in the plaquette.
Note that incorporating finite volume effects is compulsory to
see this behavior, since there are no infrared renormalons on a finite lattice.
To parameterize finite size effects, the purely perturbative OPE Eq.~(\ref{OPEpert}) was used.
The behavior seen in Fig.~\ref{n35}, although computed from
perturbative expansion coefficients, goes beyond the purely perturbative
OPE since it predicts the position of a non-perturbative object
in the Borel plane.

\section{The plaquette: OPE beyond perturbation theory}

Since in NSPT one Taylor expands in powers of $g$ before averaging
over the gauge variables, no  
mass gap is generated.
In non-perturbative Monte-Carlo (MC) lattice simulations an additional 
scale, $\lQ \sim 1/a \, e^{-2\pi/(\beta_0\al)}$,
is generated dynamically.
However, we can always tune $N$ and $\al$ such that
\be
\label{eq:scalesNP1}
\frac{1}{a} \gg \frac{1}{Na}  \gg \lQ\,.
\ee
In this small-volume situation one encounters a double expansion in powers of 
$a/(Na)$ and $a\lQ$ [or, equivalently,
$(Na)\lQ \times a/(Na)$]. The construction of the OPE
is completely analogous to that of the previous section
and one obtains\footnote{
In the last equality, we approximate the Wilson coefficients by their
perturbative expansions,
neglecting the possibility of non-perturbative contributions
associated to the hard scale $1/a$. These  
would be suppressed by factors $\sim\exp(-2\pi/\alpha)$
and, therefore, would be sub-leading relative to the gluon condensate.}
\be
\label{OPEMC}
\langle P\rangle_{\MC} =\frac{1}{Z}\left.\int\![dU_{x,\mu}]\,e^{-S[U]} P[U]
\right|_{\MC}
=
P_{\mathrm{pert}}(\al)\langle 1 \rangle
+\frac{\pi^2}{36}C_{\G}(\al)\,a^4\langle G^2 \rangle_{\MC}
+\mathcal{O}(a^6)\,.
\ee
In the last equality we have factored out the hard scale $1/a$ from the 
scales $1/(Na)$ and $\lQ$, which are encoded in
$\langle G^2\rangle_{\MC}$. Exploiting the right-most
inequality of Eq. (\ref{eq:scalesNP1}),
we can expand $\langle G^2 \rangle_{\MC}$ as follows:
\be
\langle G^2\rangle_{\MC}=\langle G^2 \rangle_{\rm soft}\left\{1+\mathcal{O}[\lQ^2 (Na)^2]\right\}
\,.
\ee
Hence, a non-perturbative
small-volume simulation
would yield the same expression as NSPT,
up to non-perturbative corrections that can be made arbitrarily
small by reducing $a$ and therefore
$Na$, keeping $N$ fixed.
In other words, $p_n^{\mathrm{NSPT}}(N)=p_n^{\MC}(N)$ up to
non-perturbative corrections.

We can also consider the limit 
\be
\label{eq:scalesNP2}
\frac{1}{a}  \gg \lQ \gg \frac{1}{Na} \,.
\ee
This is the standard situation realized in non-perturbative lattice simulations.
Again the OPE can be constructed as in the previous section, Eq. (\ref{OPEMC})
holds, and the $p_n$- and $c_n$-values are still the same.
The difference is that now 
\be
\langle G^2 \rangle_{\MC}=
\langle G^2 \rangle_{\NP}\left[1+\mathcal{O}\left(\frac{1}{\lQ^2 (Na)^2}\right)
\right]
\,,
\ee
where $\langle G^2\rangle_{\NP} \sim \lQ^4$ is the so-called
non-perturbative gluon condensate introduced
in Ref.~\cite{Vainshtein:1978wd}. From now on we will call
this quantity simply the ``gluon condensate'' $\langle G^2\rangle$. 
Nevertheless, without further qualifications, this quantity is ill defined.  

The perturbative sum and the leading nonperturbative correction in \eqref{OPEMC} are ill-defined. The reason is that the perturbative series is divergent due
to renormalons~\cite{Hooft} (for a review see \cite{Beneke:1998ui}) and other, subleading, instabilities.
This makes any determination of $\langle G^2\rangle$ ambiguous,
unless we define how to truncate or how to
approximate the perturbative series. Any reasonable definition consistent with
$\langle G^2\rangle \sim \Lambda^4$ can only be given if the asymptotic
behaviour of the perturbative series is under control.
This has only been achieved recently in Ref. \cite{Bali:2014fea},
where the perturbative expansion of the plaquette was
computed up to $\mathcal{O}(\al^{35})$. The observed
asymptotic behaviour was in full compliance with renormalon
expectations. 

Extracting the gluon condensate from the
average plaquette was
pioneered in Refs.~\cite{Di Giacomo:1981wt,Kripfganz:1981ri,DiGiacomo:1981dp,Ilgenfritz:1982yx}, and
many attempts followed during the next decades,
see, e.g., Refs.~\cite{Alles:1993dn,DiRenzo:1994sy,Ji:1995fe,DiRenzo:1995qc,Burgio:1997hc,Horsley:2001uy,Rakow:2005yn,Meurice:2006cr,Lee:2010hd,Horsley:2012ra}. Nevertheless, 
they suffered from insufficiently high perturbative orders and,
in some cases, also finite volume
effects. The failure to make a controlled contact to the asymptotic regime 
prevented a reliable lattice determination of $\langle G^2\rangle$, where one could quantitatively assess the error associated to these determinations. This problem was first solved in \cite{Bali:2014sja}. In such paper, for the first time, the perturbative sum was computed with superasymptotic accuracy for the case of 4 dimensional SU(3) gluodynamics. This allowed to obtain a reliable determination of $\langle G^2 \rangle$ that scaled as $\Lambda^4$. One issue raised was to determine to which extent such a result was independent of the scheme used for the coupling constant. The answer to this question was given within the general framework of hyperasymptotic expansions. First in \cite{Ayala:2019uaw}, where it was concluded that the error of using the superasymptotic approximation to the perturbative sum was of ${\cal O}(\sqrt{\alpha(1/a)} Z_P\Lambda^4)$, where $Z_P$ is the normalization of the leading renormalon. This error then sets the parametric precision of the determination of the gluon condensate using the superasymptotic approximation. Note that the scheme dependence of $Z_P$ and $\Lambda^4$ cancels each other. Therefore, the only remaining/leading scheme/scale dependence of the error is due to the $\sqrt{\alpha(1/a)}$ prefactor. 

To improve over the asymptotic accuracy, the first step is to regularize the perturbative sum, which we do using the Principal Value (PV) prescription. Only after regularizing the perturbative sum, the definition of the gluon condensate is unambiguous and the operator product expansion of the plaquette reads
\be
\label{OPEPV}
\langle P \rangle_{\mathrm{MC}}=
S_{\rm PV}
+\frac{\pi^2}{36}C_{\rm G}(\al)\,a^4\langle G^2  \rangle_{\rm PV}+\mathcal{O}\left((a\Lambda)^6\right)\,.
\ee
This expression is, in practice, formal, as the exact expression of $S_{\rm PV}$ is not known. This would require the exact knowledge of the Borel transform of the perturbative sum. Nevertheless, it is possible to obtain an approximate expression of it with a known parametric control of the error using its hyperasymptotic expansion. The accuracy of this expansion is limited from the information we get from perturbation theory. For the case at hand, we have  
\be
\label{eq:SPV}
S_{\rm PV}=
\sum_{n=0}^{N_P}p_n\alpha^{n+1}+\Omega_{G^2}+
\sum_{n=N_P+1}^{N'}[p_n-p_n^{\rm (as)}]\alpha^{n+1}
+\cdots\,,
\ee
where $N'$ is the maximal order in perturbation theory that is included in the perturbative expansion. Within the hyperasymptotic counting, approximating $S_{\rm PV}$ by $S_{\rm P}\equiv 
\sum_{n=0}^{N_P}p_n\alpha^{n+1}$, the first term in \eqref{eq:SPV}, corresponds to the superasymptotic approximation, which we label as $(0,N_P)$. Adding $\Omega_{G^2}$ to the superasymptotic approximation corresponds to (4,0) precision in the hyperasymptotic approximation and adding the last term corresponds to $(4,N')$ precision\footnote{The labeling (D,N) in general is defined in Refs. \cite{Ayala:2019hkn,Ayala:2019lak}.}. 

In \eqref{eq:SPV}, we take
\be
\label{eq:NP}
N_P=4\frac{2\pi}{\beta_0\al(1/a)}\left(1-c\al(1/a)\right)
\,,
\ee
as the order at which we truncate the perturbative expansion to reach the superasymptotic approximation. By default, we will take the smallest positive value of $c$ that yields an integer value for $N_P$, but we also explore the dependence of the result on $c$. Note that the value of $N_P$ used in Eq. (\ref{eq:NP}) is slightly different from the value used in \cite{Bali:2014sja} to truncate the perturbative expansion with superasymptotic accuracy. In that reference, such number was named $n_0$ and was determined numerically. We will ellaborate on this difference later. 

$\Omega_{G^2}$ is the terminant associated with the leading renormalon of the plaquette. It reads \cite{Ayala:2020pxq}

\begin{align}
	\label{OmegaExp}
	\Omega_{G^2}=&
	\sqrt{\al(1/a)}K^{(P)}
	e^{-\frac{8\pi}{\beta_0 \al(1/a)}}
	\left(\frac{\beta_0\al(1/a)}{4\pi}\right)^{-4b}
	\bigg(
		1
		+\bar K_{1}^{(P)}\al(1/a)
		\nonumber
		\\
		&+\bar K_{2}^{(P)}\al^2(1/a)
		+\mathcal{O}(\alpha^3(1/a))\bigg)
\,,
\end{align}
where 
\begin{align}
	&K^{(P)}=
	\frac{-Z_P}{\Gamma(1+4b)}
	2^{2+4b}\pi\beta_0^{-1/2}
	\left(-\eta_c+\frac{1}{3}\right)
\,,
	\\
	&\bar K_{1}^{(P)}=
	\frac{\beta_0/(4\pi)}{-\eta_c+\frac{1}{3}}
	\bigg[
		-4bb_1\left(\frac{1}{2}\eta_c+\frac{1}{3}\right)
		-\frac{1}{12}\eta_c^3
		+\frac{1}{24}\eta_c
		-\frac{1}{1080}\bigg]
\,,	\\
	&K_{1}^{(P)}=\bar K_{1}^{(P)}-\frac{2b \beta_0 s_1}{\pi}
\,,	\\
	&\bar K_{2}^{(P)}=
	\frac{\beta_0^2/(4\pi)^2}{-\eta_c+\frac{1}{3}}
	\bigg[
		-4 w_2(4b -1)b\left(\frac{1}{4}\eta_c+\frac{5}{12}\right)
		\nonumber
		\\
		&\qquad
		+4b_1 b\left(
			-\frac{1}{24}\eta_c^3
			-\frac{1}{8}\eta_c^2
			-\frac{5}{48}\eta_c
			-\frac{23}{1080}\right)
		-\frac{1}{160}\eta_c^5
		-\frac{1}{96}\eta_c^4
		+\frac{1}{144}\eta_c^3
		\nonumber
		\\
		&\qquad
		+\frac{1}{96}\eta_c^2
		-\frac{1}{640}\eta_c
		-\frac{25}{24192}\bigg]
\,,	\\
	&K_{2}^{(P)}=
	\frac{1}{8\pi^2}
	\big(
		8\pi^2\bar K_{2}^{(P)}
		-16b\pi s_1\beta_0\bar K_{1}^{(P)}
		+16b^2s_1^2\beta_0^2
		+8b^2s_2\beta_0^2\big)
\,,
\end{align}
where $\eta_c\equiv-4b+\frac{8\pi}{\beta_0}c-1$.

The value of $Z_P$ was determined approximately (for $n_f=0$) in \cite{Bali:2014fea}:
\be
Z_P=(42\pm 17)\times 10^4
\,.
\ee
Its error gives the major source of uncertainty in the determination of $\Omega_{G^2}$, of the order of 40\%. The other source of error is due to the fact that only approximate expressions are  available for $\Omega_{G^2}$, as we do not know the complete set of coefficients of the beta function in the lattice scheme. Nevertheless, we can study the convergence pattern of the weak-coupling expansion. Equation \eqref{OmegaExp} yields a nicely convergent series with a controlled scheme dependence, as the weak coupling expansion is organized in terms of a single parameter: $\alpha$. The error associated with truncating the expansion in \eqref{OmegaExp} is estimated by observing the convergent pattern of the LO, NLO and NNLO results. From LO to NLO, in the worst cases, the differences are close but below 50\%, and from NLO to NNLO, the differences are below 10\%. One could then expect the NNNLO contribution to be at the level of few percent, which can be neglected all together in comparison with the $\sim 40\%$ error associated to $Z_P$.

\section{Determination of the gluon condensate} 

We now review the determination of the gluon condensate in \cite{Bali:2014sja,Ayala:2020pxq}. We determine the gluon condensate from the following equation: 
\be
\label{eq:Fit}
\langle G^2 \rangle_{\rm PV} = \frac{36 C_G^{-1}}{\pi^2 a^4}
\left[ \langle P \rangle_{\rm MC} -S_{\rm PV}
\right]
\,.
\ee
If $S_{\rm PV}$ and $\langle P \rangle_{\rm MC}$ were known exactly, this equality is expected to hold up to corrections of ${\cal O}(a^2\Lambda^2)$. Nevertheless, neither $S_{\rm PV}$ nor $\langle P \rangle_{\rm MC}$ are known exactly. On top of that, $C_G^{-1}$ and the relation between $a$ and $\beta$ are also known in an approximated way. We now discuss how to determine them and their associated individual errors. 

\begin{figure}
\begin{center}
\resizebox{0.75\columnwidth}{!}{  \includegraphics{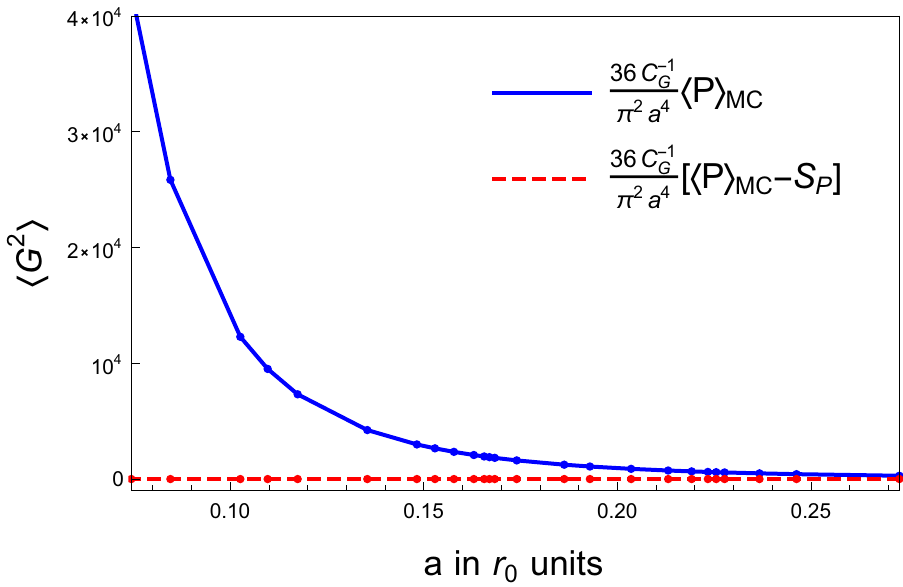} }
\end{center}
\caption{$ \frac{36 C_G^{-1}}{\pi^2 a^4} \langle P \rangle_{\rm MC}$ (continuous blue line) and $ \frac{36 C_G^{-1}}{\pi^2 a^4}\left[ \langle P \rangle_{\rm MC} -S_{P} \right]$ (dashed red line). The second line is basically indistinguishable with respect to zero with the scale resolution of this plot. The statistical errors are smaller than the size of the points. From \cite{Ayala:2020pxq}.}
\label{Fig:cancellation}
\end{figure}

The MC data is taken from \cite{Boyd:1996bx}, restricting to the more precise
$N=32$ data and, to keep finite volume effects under control,
to $\beta\leq 6.65$. We also 
limit ourselves to $\beta\geq 5.8$ to avoid large
$\mathcal{O}(a^2)$ corrections. 
At very large $\beta$-values, obtaining meaningful results becomes challenging
numerically: the individual errors both of $\langle P\rangle_{\MC}(\al)$
and of $S_{\rm PV}(\al)$ somewhat decrease with increasing $\beta$.
However, there is a very strong cancellation between these two terms, in
particular at large $\beta$-values, since this difference
decreases with $a^{-4}
\sim\Lambda_{\latt}^4\exp(16\pi^2\beta/33)$ on dimensional
grounds, while $\langle P\rangle_{\MC}$ depends only logarithmically
on $a$. We illustrate this cancellation in Fig. \ref{Fig:cancellation}. 

Equation~(\ref{eq:betafun}) is not accurate enough in the lattice scheme
for the available $\beta$-values.
Instead, the
phenomenological parametrization of Ref.~\cite{Necco:2001xg}
($x=\beta-6$)
\begin{align}
\label{eq:Necco}
a=r_0\exp\left(-1.6804-1.7331x+0.7849x^2-0.4428x^3\right)\,,
\end{align}
obtained by
interpolating non-perturbative lattice simulation results is used.
Equation~(\ref{eq:Necco}) was reported to be valid within an accuracy varying
from 0.5\% up to 1\% in the
range~\cite{Necco:2001xg} $5.7\leq\beta\leq 6.92$, which includes the range $\beta \in [5.8,6.65]$ used here. 

For the inverse Wilson coefficient
\begin{align}
C^{-1}_{\rm G}(\al)=
\label{CPinverse}
-\frac{2\pi\beta(\al)}{\beta_0\al^2}
=1+\frac{\beta_1}{\beta_0}\frac{\al}{4\pi}
+\frac{\beta_2}{\beta_0}\left(\frac{\al}{4\pi}\right)^2
+\frac{\beta_3}{\beta_0}\left(\frac{\al}{4\pi}\right)^3
+{\cal O}(\al^4)
\,,
\end{align}
the corrections to
$C_{\rm G}=1$ are small. However, the $\mathcal{O}(\alpha^2)$ and
$\mathcal{O}(\alpha^3)$ terms are of similar sizes. We will
account for this uncertainty in our error budget.

\medskip

\begin{figure}
\centering
\resizebox{0.75\columnwidth}{!}{  \includegraphics{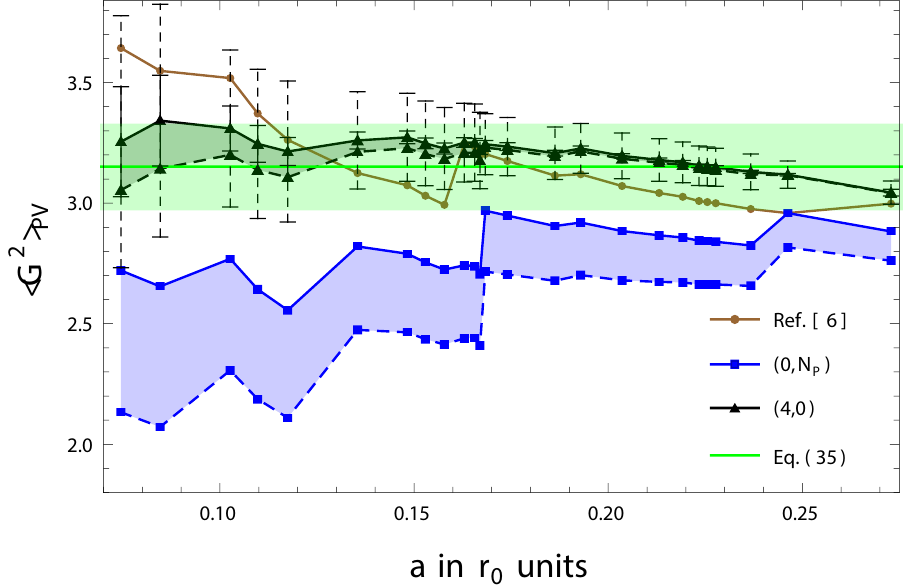} }
\caption{Gluon condensate with superasymptotic approximation $(0,N_P)$ and with hyperasymptotic accuracy $(4,0)$. In both cases, for each corresponding $\beta$, we show the value obtained for the gluon condensate with the values of $N_P$ using the smallest positive (upper line) and negative (lower line) value of $c$ that yields an integer value of $N_P$. For the hyperasymptotic approximation with $c$ positive we also show the statistical errors of the MC determination of the plaquette (inner error) and its combination in quadrature with the statistical error of the partial sum (outer error). We also show the superasymptotic approximation obtained in \cite{Bali:2014sja} (Ref. [28] in the original plot) truncating at the minimal term determined numerically. The horizontal green band and its central value are the final prediction, and the associated error, for the gluon condensate displayed in \eqref{G2final}. From \cite{Ayala:2020pxq}.}
\label{Fig:G2}
\end{figure}

We now turn to $S_{\rm PV}(\al)$. It is computed using the hyperasymptotic expansion. This introduces a parametric error according to the order we truncate this expansion. On top of that, the coefficients $p_n$, obtained in Ref.~\cite{Bali:2014fea}, are not known exactly. They carry statistical errors,
and successive orders are correlated. Using the covariance matrix, also
obtained in Ref.~\cite{Bali:2014fea}, the statistical error of $S_P(\al)$ can
be calculated. In that reference, coefficients $p_n(N)$ were first
computed on finite volumes of $N^4$ sites and subsequently extrapolated
to their infinite volume limits $p_n$. This extrapolation
is subject to parametric uncertainties that need to be
estimated. In Ref.~\cite{Bali:2014fea} and later in \cite{Ayala:2020pxq} the differences between determinations
using $N\geq \nu$ points for $\nu=9$ (the central values)
and $\nu=7$ were added as systematic errors to the statistical errors.  We emphasize though, that the order the perturbative series was truncated, $N_P$, is different in each case. The difference between both determinations gives an estimate of the parametric error of the determination of $S_{\rm PV}(\al)$ by using the superasymptotic approximation $S_P$. The magnitude of $\Omega_{G^2}$ gives an alternative estimate of the error associated with the truncation of the hyperasymptotic approximation. It is also interesting to see the magnitude of changing $N_P$ by one unit by fine tunning $c$ from the smallest positive value that yields an integer value of $N_P$ to the smallest (in modulus) negative value that yields an integer value of $N_P$. Typically this yields slightly smaller errors. We illustrate this discussion in Fig. \ref{Fig:G2}. All these error estimates scale with the parametric uncertainty predicted by theory $\sim {\cal O}(e^{-4\frac{2\pi}{\beta_0 \al(1/a)}}) \sim {\cal O}(a^4\Lambda^4)$ times $\sqrt{\alpha}$ (see the discussion in \cite{Ayala:2019hkn,Ayala:2019lak}).

%\begin{center}
%\end{center}

If we increase the accuracy of the hyperasymptotic expansion by adding the terminant $\Omega_{G^2}$ to the superasymptotic approximation, the parametric error decreases, and the accuracy reached is (4,0) (note that the statistical error does not change). With this accuracy, the parametric error is $\sim {\cal O}(e^{-4\frac{2\pi}{\beta_0 \al(1/a)}(1+\log(3/2))}) \sim  {\cal O}((a\Lambda)^{4(1+\log(3/2))})$ (see the discussion in \cite{Ayala:2019hkn,Ayala:2019lak}). 
Note that $4\log(3/2) \simeq 1.6 <2$. Therefore, these effects are parametrically more important than the next nonperturbative power corrections. Compared with the typical size of the terminant $\Omega_{G^2}$, these effects are suppressed by a factor of order $\sim {\cal O}((a\Lambda)^{4\log(3/2))})$. In the energy range we do the fits, this yields suppression factors in the range $((a\Lambda_{\MS})^{4\log(3/2)}) \in (0.007
%0.0535363
,
%0.00650847
0.05)$, where we have taken $\Lambda=\Lambda_{\MS}$ to be more conservative. This discussion can be affected by powers of $\alpha$. It is expected that there is an extra suppression factor of $\alpha^{3/2}$ (as $\sqrt{\al}$ is already included in the terminants the real suppression factor would be of order $\alpha$). Depending on the scheme, the size of this extra factor is different. In any case, they go in the direction to make the estimate of the error smaller. We will not dwell further in this discussion of the parametric error of the (4,0) hyperasymptotic accuracy, because we only approximately know $\Omega_{G^2}$ and its error will hide the signal of these ${\cal O}((a\Lambda)^{4(1+\log(3/2))})$ effects. For $\Omega_{G^2}$ we use the analytic expression in \eqref{OmegaExp} truncated at ${\cal O}(\al^2)$.
The error of this expression comes from $Z_P$, and from the truncation of the weak coupling expansion of the terminant. The largest source of error comes from $Z_P$. Due to its size, this error overwhelms  the parametric error associated to higher-order terms in the hyperasymptotic expansion. 

Irrespective of the discussion of the error of the (4,0) accuracy, it is nice to see that adding the terminant to the superasymptotic expression makes the jumps that we had with the superasymptotic approximation disappear. Adding the terminant also makes the resulting curve flatter. The dependence in $N_P$ (or in other words $c$) gets much milder too. We illustrate all this in Fig. \ref{Fig:G2}.

\medskip

In principle, we know perturbation theory to orders high enough to include the last term written in \eqref{eq:SPV} and reach $(4,N')$ accuracy. Nevertheless, we find that the errors of $p_n$ for large $n$ hide the signal, and it is not possible to improve the prediction. The optimal value is given below in \eqref{G2final}. For further details in the error analysis see Ref. \cite{Ayala:2020pxq}.

\section{Conclusions}

For the first time ever, perturbative expansions at orders
where the asymptotic regime is reached were obtained \cite{Bali:2014fea} and subtracted
from non-perturbative Monte Carlo data with superasymptotic \cite{Bali:2014sja} and hyperasymptotic accuracy \cite{Ayala:2020pxq}. The most accurate value was obtained in this last reference:
\be
\label{G2final}
\langle G^2 \rangle_{\rm PV} (n_f=0)=3.15(18)\, r_0^{-4}
\,.
\ee
We emphasize that this result is independent of the scale and renormalization scheme used for the coupling constant. Even if the computation was made in the lattice scheme, the result is the same in the $\MS$ scheme within the accuracy of the computation. The limiting factor for improving the determination of the gluon condensate in pure gluodynamics is the error of perturbation theory. All systematic sources of error have its origin in the errors of perturbation theory (even what we call statistical errors of \eqref{eq:Fit} are dominated by the statistical errors of the coefficients $p_n$). More precise values of these perturbative coefficients, and its knowledge to higher orders, would yield a more precise determination of the normalization of the renormalons, $Z_P$, and would allow working with hyperasymptotic accuracy $(4,N')$. Nowadays, if we try to reach this accuracy, we find that the error of the coefficients are too large to get accurate results. The situation with active light quarks is in an early stage but starts to be promising. The coefficients of the perturbative coefficients have been computed at finite volume in \cite{DelDebbio:2018ftu} for QCD with two massless fermions.  More data at different volumes, and the infinite volume extrapolation of these coefficients, would then allow to give a determination of the gluon condensate in QCD with two massless fermions. 

\begin{figure}[htb]
\centering
\resizebox{0.75\columnwidth}{!}{  \includegraphics{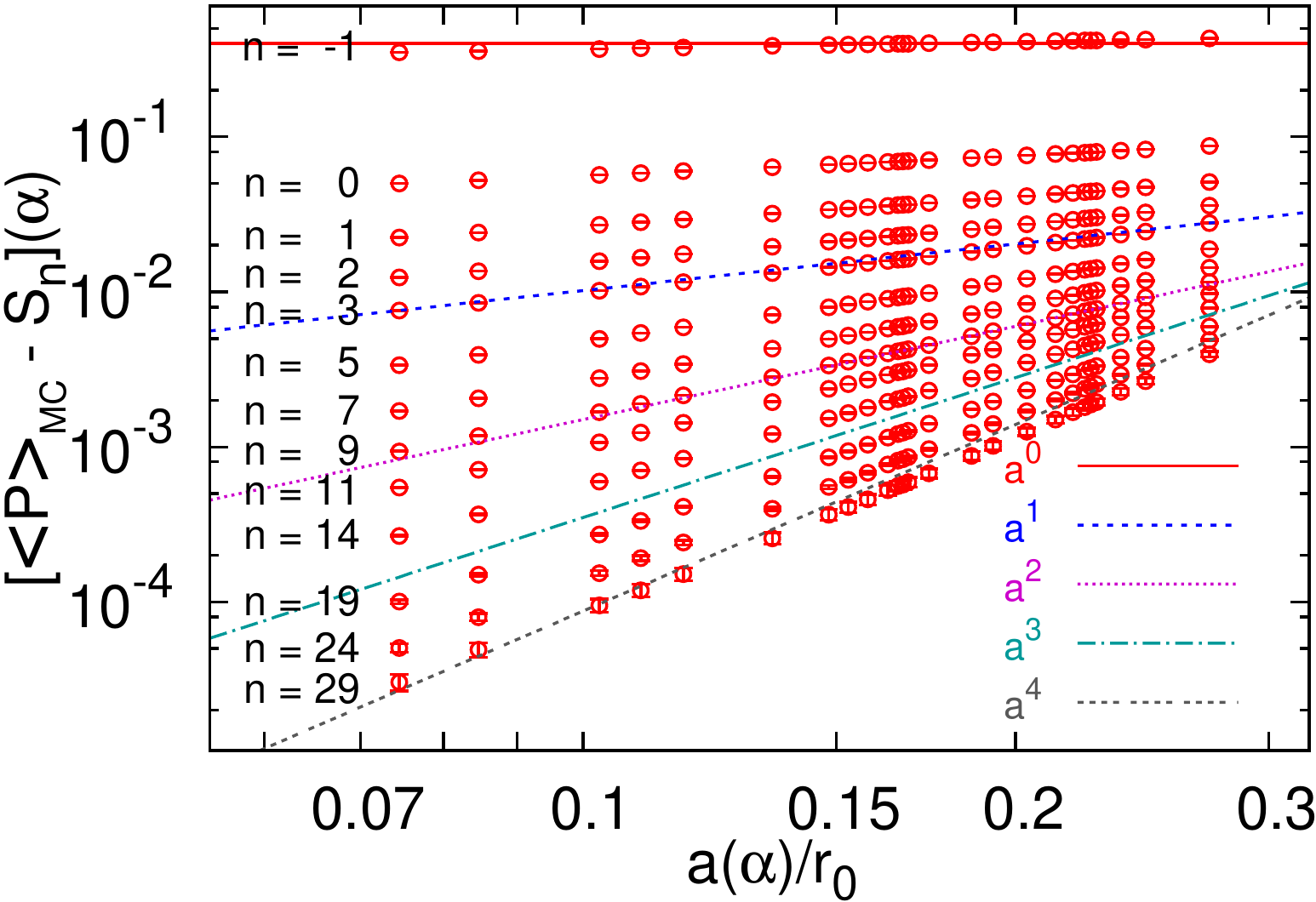} }
\caption{Differences $\langle P\rangle_{\MC}(\al)-S_n(\al)$ (where $S_n(\al) \equiv \sum_{s=0}^n p_s\alpha^{s+1}$)
between MC data and sums truncated
at orders $\al^{n+1}$ ($S_{-1}=0$)
vs.\ $a(\al)/r_0$. The lines $\propto a^j$ are drawn
to guide the eye. From \cite{Bali:2014sja}.
\label{fig:a2}}
\end{figure}

Overall, the OPE beyond perturbation theory has been validated. The scaling of the nonperturbative effects with the lattice spacing
confirms the dimension $d=4$. Dimension $d<4$ slopes appear only
when subtracting the perturbative series truncated at fixed
pre-asymptotic orders. Therefore, these lower dimensional ``condensates'' discussed in Ref.~\cite{Chetyrkin:1998yr} or in Ref.~\cite{Burgio:1997hc} are nothing but approximate parametrizations of unaccounted perturbative effects, i.e., of the short-distance behavior. These will be observable-dependent,
unlike the non-perturbative gluon condensate. Such simplified
parametrizations of the perturbative terms introduce unquantifiable errors and, therefore, are of limited phenomenological use. As illustrated in Fig.~\ref{fig:a2}, even the 
effective dimension of such a ``condensate'' varies when
truncating a perturbative series at different orders. In
Refs.~\cite{Gubarev:2000nz,RuizArriola:2006gq,Andreev:2006vy} various analyses, based on models such as string/gauge duality 
or Regge models, have been made claiming the existence of non-perturbative dimension two corrections. Our results 
strongly suggest that there may be flaws in these derivations.

The accurate value obtained for
the gluon condensate, Eq.~(\ref{G2final}), is of a similar size as the intrinsic
difference between
(reasonable) subtraction prescriptions (see the discussion in \cite{Bali:2014sja}). This result contradicts the 
implicit assumption of sum rule analyses that the renormalon ambiguity is 
much smaller than leading non-perturbative corrections.
The value of the gluon condensate obtained with sum rules
can vary significantly due to this intrinsic
ambiguity if determined using different
prescriptions or truncating at different orders in perturbation theory.
Clearly, the impact of this, e.g., on determinations
of $\alpha_s$ from $\tau$-decays or from lattice simulations needs to be
assessed carefully.

We finally mention that the nonzero value of $\langle G^2 \rangle_{\rm PV}$ shows that the PV regularization of the perturbative sum, even if computed exactly, would differ from the Montecarlo simulation of the plaquette by a term of ${\cal O}(a^4\Lambda^4)$. This may affect the conjecture that the resummation technique of the perturbative expansion proposed in \cite{Caprini:2020lff} for the Adler function would not need such nonperturbative corrections. This should be further investigated.

\medskip
 
{\bf Acknowledgments}\\
\noindent
I thank C. Ayala, G.S. Bali, C. Bauer, X. Lobregat for collaboration in the work reviewed here. 
This work was supported in part by the Spanish grants FPA2017-86989-P and SEV-2016-0588 from the ministerio de Ciencia, Innovaci\'on y Universidades, and the grant 2017SGR1069 from the Generalitat de Catalunya. This project has received funding from the European Union's Horizon 2020 research and innovation programme under grant agreement No 824093. IFAE is partially funded by the CERCA program of the Generalitat de Catalunya.

\end{document}